\documentclass[twocolumn,twoside,slac_two]{revtex4}
\usepackage{graphicx}
\usepackage{fancyhdr}
\newcommand{\chan}{{\em Chandra}}
\newcommand{\xmm}{{\em XMM-Newton}}

\newcommand{\vlt}{{\em VLT}}

\newcommand{\hst}{{\em HST}}

\newcommand{\fors}{{\em FORS2}}

\begin{document}

\title{VLT observations of Fermi pulsars}

%

\author{R. P. Mignani}
\affiliation{Mullard Space Science Laboratory, University College London, Holmbury St. Mary, Dorking, Surrey, RH5 6NT, UK \\
Institute of Astronomy, University of Zielona G\'ora, Lubuska 2, 65-265, Zielona G\'ora, Poland}

\author{A. Shearer}
\affiliation{Centre for Astronomy, National University of Ireland, Newcastle Road, Galway, Ireland}

\author{A. De Luca}
\affiliation{IUSS - Istituto Universitario di Studi Superiori, viale Lungo Ticino Sforza, 56, 27100, Pavia, Italy \\
INAF - Istituto di Astrofisica Spaziale e Fisica Cosmica Milano, via E. Bassini 15, 20133, Milano, Italy \\
INFN - Istituto Nazionale di Fisica Nucleare, sezione di Pavia, via A. Bassi 6, 27100, Pavia, Italy}
   
\author{P. Moran}
\affiliation{Centre for Astronomy, National University of Ireland, Newcastle Road, Galway, Ireland}

\author{S. Collins}
\affiliation{Centre for Astronomy, National University of Ireland, Newcastle Road, Galway, Ireland}

\author{M. Marelli}
\affiliation{INAF - Istituto di Astrofisica Spaziale e Fisica Cosmica Milano, via E. Bassini 15, 20133, Milano, Italy \\
Universit\'a degli Studi dell' Insubria, Via Ravasi 2, 21100, Varese, Italy}
   
\begin{abstract}
Many energetic $\gamma$-ray pulsars discovered by Fermi are promising candidates for optical follow-ups. We present the results of the first deep optical observations of the two Vela-like Fermi pulsars PSR\, J1357$-$6429 and PSR\, J1048$-$5832 performed with the VLT. However, they have not been detected down to V$\sim$27 and V$\sim$27.6, respectively ($3 \sigma$). These upper limits suggest an efficiency in converting spin-down power into optical luminosity $< 7 \times 10^{-7}$ and $<6 \times 10^{-6}$, respectively, lower than the Crab pulsar and, possibly, more compatible with the spin-down age of these two pulsars.
\end{abstract}

\maketitle

\thispagestyle{fancy}

\section{INTRODUCTION}

Optical observations of pulsars are crucial to study the intrinsic properties of neutron stars, from the structure and composition of the interior, to the properties and geometry of the magnetosphere. Historically, X-ray and $\gamma$-ray observations have paved the way to the pulsar optical identifications, with all the presently identified pulsars also detected at high energies (Mignani 2011). In particular, 5 out of
the 7 $\gamma$-ray pulsars detected by NASAÕs {\em CGRO} satellite have also been detected in the optical, suggesting that $\gamma$-ray detections highlight promising candidates for optical observations, since the emission at both energies seems to correlate with the strength of the magnetic field at the light cylinder (e.g., Shearer \& Golden 2001). The launch of the {\em Fermi} Gamma-ray Space Telescope opened new perspectives
with the detection of over 80 $\gamma$-ray pulsars (Abdo et al.\ 2011). For most {\em Fermi} pulsars, however, no deep optical observations have been carried out so far. In the Northern hemisphere, an exploratory survey  with 2.5m/4m telescopes at the La Palma Observatory has been performed (Shearer et al., these proceedings), while in the Southern hemisphere a survey with the {\em VLT} is in progress (see Mignani et al. 2011a).  Here, we present the search for the optical counterparts of two {\em Fermi} pulsars (Mignani et al.\ 2011b). PSR\, J1357$-$6429 is a very young (7.31 kyr old) pulsar (P = 166.1 ms), at a distance of $\sim$2.5 kpc, with a rotational energy loss rate $\dot{E} \sim 3 \times 10^{36}$ erg s$^{-1}$. It has been detected as a $\gamma$-ray pulsar both by {\em AGILE} (Pellizzoni et al.\ 2009)  and {\em Fermi} (Lemoine-Goumard et al.\ 2011), while  its associated pulsar wind nebula (PWN) has been detected at TeV energies by HESS (Abramowski et al. 2011). PSR\, J1048$-$5832 is a 20.3 kyr old, Vela-like, pulsars (P=123.6 ms), tentatively detected by the {\em CGRO} (Kaspi et al.\ 2000) and now confirmed by {\em Fermi} (Abdo et al.\ 2009), at a distance of $\sim$2.7 kpc and with a $\dot{E} \sim 2 \times 10^{36}$ erg s$^{-1}$ comparable to that of PSR\, J1357$-$6429. Both pulsars are also detected in X-rays by both {\em XMM-Newton} and {\em Chandra} (Esposito et al.\ 2007; Lemoine-Goumard et al.\ 2011; Chang et al.\  2011; Gonzalez et al.\ 2006) . No deep optical observations of these two pulsars have been ever reported. We used multi-band {\em VLT} images  available in the ESO public archive to search for their optical emission

\section{OBSERVATIONS}

\begin{figure*}[t]
\includegraphics[height=7cm]{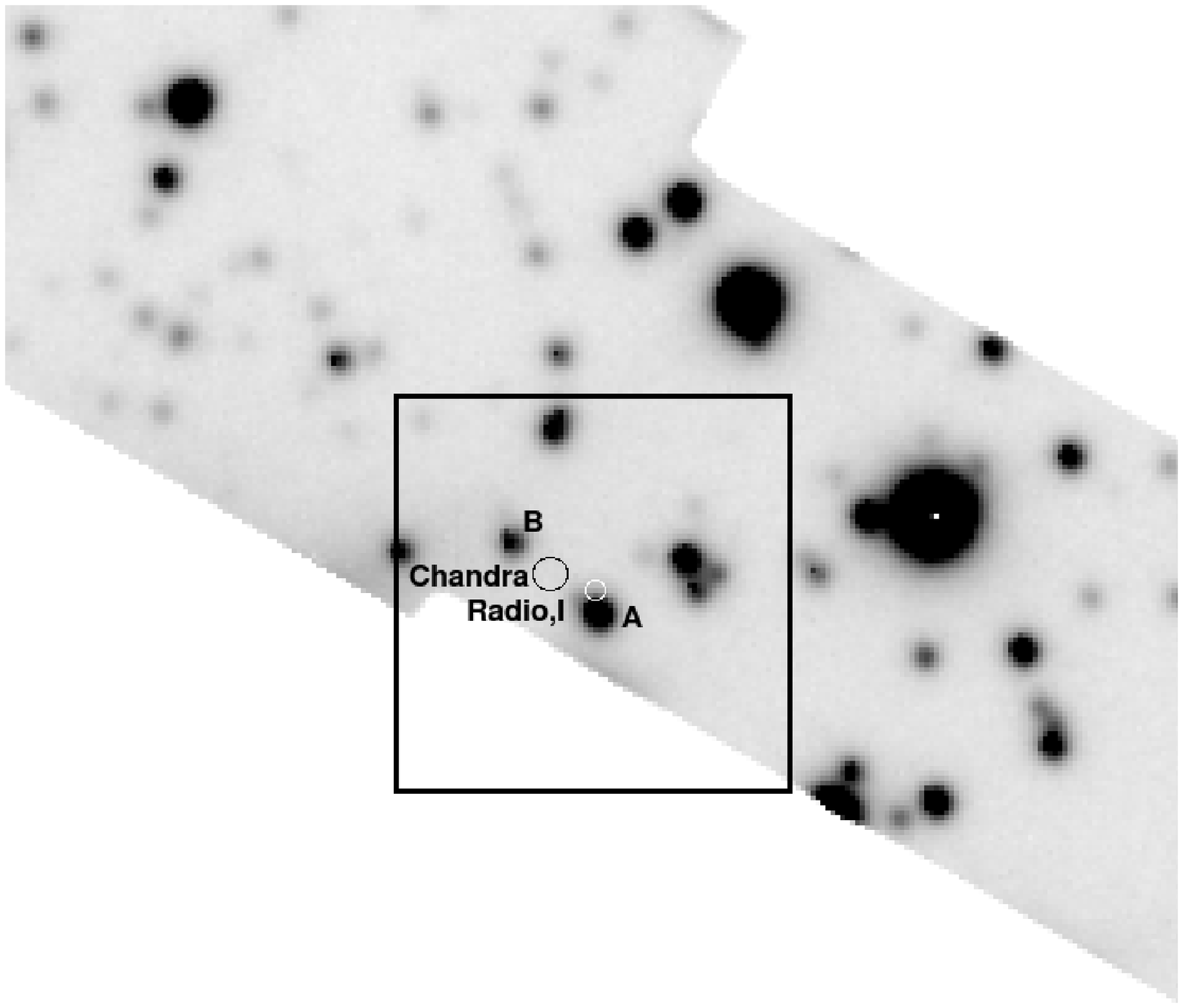}
\includegraphics[height=7cm]{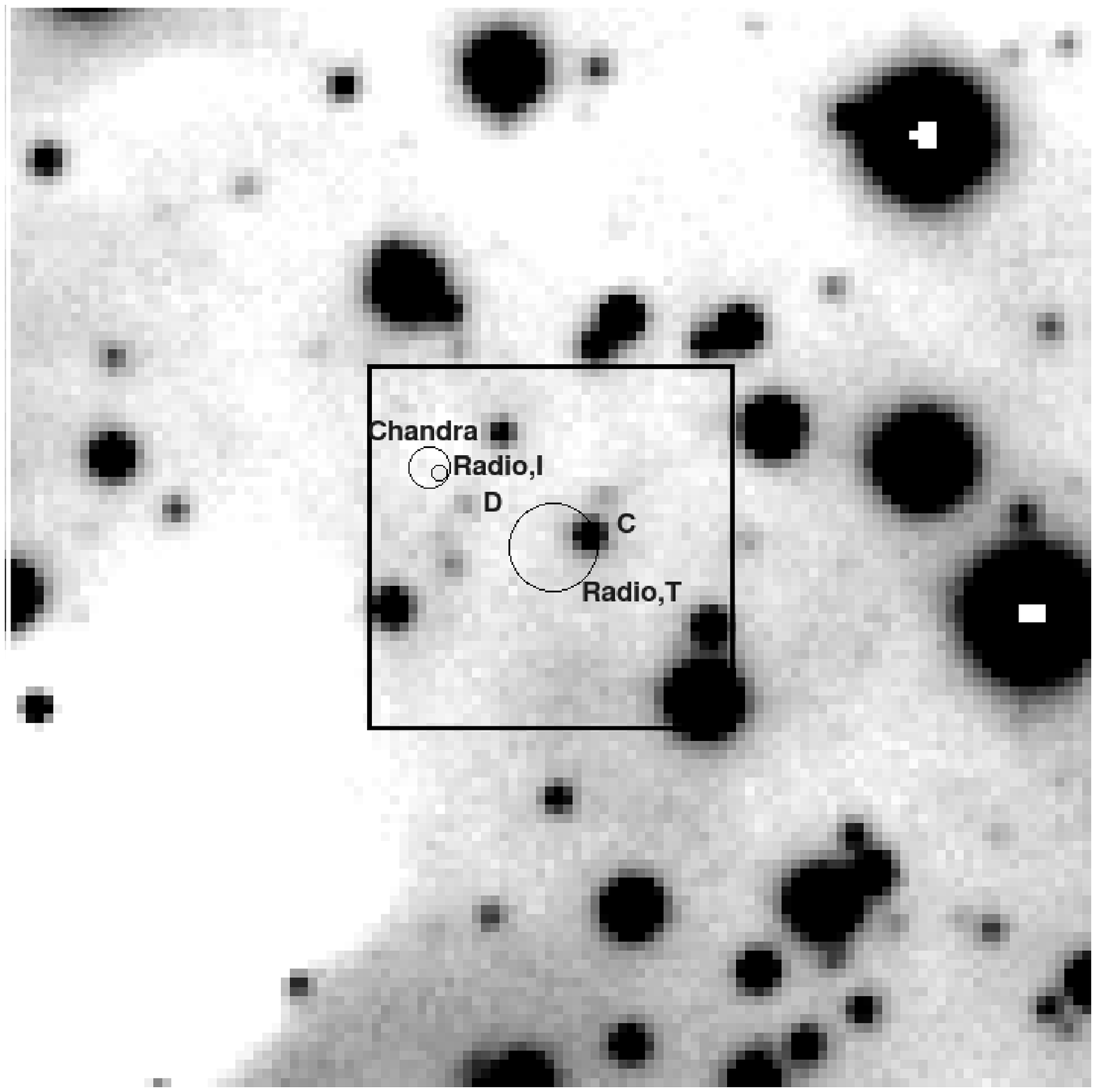}
  \caption{ {\em VLT}/FORS2  30$\times$30 arcsec image cutouts  of PSR\, J1357$-$6429 (left) and PSR\, J1048$-$5832 (right) taken through the V filter. North to the top, East to the left. in size. The circles marks the pulsar radio (T=timing; I=interferometry) and {\em Chandra} positions.   }
\end{figure*}

Optical images of the PSR\, J1357$-$6429 and PSR\, J1048$-$5832 fields were obtained with the {\em VLT} between April 2009 and February 2010. Observations were performed in service mode with FORS2, a multi-mode camera for imaging and long-slit/multi-object spectroscopy (MOS). FORS2 was equipped with its red-sensitive MIT detector, a mosaic of two 2k?4k CCDs optimised for wavelengths longer than 6000 \AA. In its standard resolution mode, the detector has a pixel size of 0.25 arcsec ($2\times2$ binning) which corresponds to a fieldÐofÐview of $8.3 \times 8.3$ arcmin over the CCD mosaic. Observations were performed with the standard low gain, fast read-out mode and in high-resolution mode (0.125 arcsec/pixel) for PSR\, J1357$-$6429 (V,R,I bands) and in standard resolution mode (0.25"/pixel) for PSR\, J1048$-$5832 (V band).  Bright stars close to the PSR\, J1357$-$6429 position were masked using the MOS slitlets. To allow for cosmic ray removal and minimise saturation of bright stars in the field, sequences of short exposures (from 200 to 750 s) were obtained per each target and per each filter. The total integration time was 14700 s (V), 8800 s (R), and 1600 s (I) for PSR\, J1357$-$6429 and 24000 s (V) for PSR\, J1048$-$5832. Exposures were taken in dark time and photometric conditions, close to the zenith (airmass$<$1.3), and sub-arcsec image quality.
We reduced the data through standard packages in {\em IRAF} for bias subtraction, and flat-field correction using the closest-in-time bias and twilight flat-fields frames available in the ESO archive. Per each band, we aligned and average-stacked the reduced science images using the IRAF task {\tt drizzle} applying a $3 \sigma$ filter on the single pixel average to filter out residual hot and cold pixels and cosmic ray hits. We applied the photometric calibration by using the extinction-corrected night zero points computed by the FORS2 pipeline and available through the instrument data quality control database (www.eso.org/qc).To register the pulsar positions on the FORS2 frames as precisely as possible, we re-computed their astrometric solution. We measured the star centroids through Gaussian fitting using the {\tt GAIA} tool and used the code {\tt ASTROM}  to compute the pixel-to-sky coordinate
transformation. For both pulsar fields, we  estimate that the overall ($1 \sigma$) uncertainty of our FORS2 astrometry is $\sim$ 0.2 arcsec.

\section{Results}

As a reference to compute the pulsar positions on the FORS2 frames we used their most recently published radio and {\em Chandra} coordinates (see Mignani et al.\ 2011b). The radio and Chandra positions of PSR\, J1357$-$6429 and PSR\, J1048$-$5832 are shown in Fig. 1. For the
former the {\em Chandra} position, averaged over multi-epoch consistent measurements, differs by $1.2 \pm 0.4$ arcsec from the radio interferometry one, an amount which can be explained by the $\sim$7.2 yr time span between the epochs for a pulsar transverse velocity of $2100\pm700$ km s$^{-1}$. For the latter, the difference between the {\em Chandra} and the radio timing positions amounts to $4.1\pm1.3$ arcsec and cannot be explained by the pulsar proper motion between the two epochs ($\sim$ 5.4 yrs apart). However, the {\em Chandra} position is consistent with the radio-interferometry one, suggesting that the radio timing position was affected by timing noise. Since the multi-epoch {\em Chandra} coordinates appear more reliable, in the following we assume them as a reference, although we conservatively  evaluate possible candidates at the radio positions.

The radio position of PSR\, J1357$-$6429 (Fig. 1, left) falls close (0.65 arcsec) to a relatively bright field object (object A; V=21.8$\pm$0.06). We verified whether object A can be considered a possible counterpart to the pulsar. The chance coincidence probability which turns out to be $\sim$0.01, i.e. not statistically compelling yet. If it were its counterpart, the flux of object A would imply an optical emission efficiency for PSR\, J1357$-$6429 up to 10 times larger than the Crab pulsar and up to $\sim$5600 times larger than the Vela pulsar, to which it is more similar in its spin parameters. Thus, it is unlikely that object A is the pulsar counterpart.  We also compared object AÕs colours (V=21.80$\pm$0.06; R=20.65$\pm$0.03; I=19.75$\pm$0.03) with those of field stars and we found that it has no peculiar colours which may suggest its association with the pulsar. For instance, for a flat power-law spectrum, like that of the Vela pulsar, we would expect a (V-R)$\sim$0.2 and a (V-I)$\sim$0.1. Thus, we conclude that object A is unrelated to PSR\, J1357$-$6429. The same arguments also rule out object B (V=23.05$\pm$0.13; R=21.86$\pm$0.07; I=20.5$\pm$0.05) as a candidate counterpart. We found a marginal evidence of a flux enhancement over the background at the edge of the \chan\ error circle, which might come from the presence of a faint source ($I\approx 24.6$). However, the local crowding, with two relatively bright stars within a radius of $\sim 1.5$ arcsec from the \chan\ position, and the lack of detections in the V and R-bands, make it problematic to determine whether such an enhancement come from a background fluctuation, perhaps produced by the PSF wings of the two stars, or it is associated with a real source and hence to a putative pulsar counterpart.  

The radio-timing position of PSR\, J1048-5832 (Fig. 1, right) falls close to a V$\sim$24 object (C). Again, the chance coincidence probability is $\sim$0.04, i.e. certainly not low enough to statistically claim an association. This is also ruled out by the pulsarÕs energetics. The flux of object C would imply an optical emission efficiency for PSR\, J1048$-$5832 up to 60 times larger than the Crab pulsar and up to $\sim$200000 times larger than the Vela pulsar. Thus, it is extremely unlikely that object C is associated with PSR\, J1048$-$5832. Then, we  searched for a possible counterpart at the pulsar \chan\  position. No object is detected within, or close to, the  \chan\ (0.55 arcsec) error circle apart from a faint object (Star D; $V\sim 26.7$) visible $\sim 1.6$ arcsec southwest of it. However, the offset is about  three times the $1	 \sigma$  uncertainty on the pulsar position. Thus, we deem the association unlikely both on the basis of the loose positional coincidence and on statistical grounds, with a chance coincidence probability of $P\sim 0.08$.   Moreover, the lack of colour information makes it impossible to constrain the object's nature. 

No other possible counterparts are detected for both PSR\, J1357$-$6429 and PSR\, J1048$-$5832. We conclude that they are unidentified in the {\em VLT} images and we computed the $3 \sigma$ upper limits on their flux. For PSR\, J1357$-$6429 we derive V$>$27.3, R$>$26.8, and I$>$24.6, while for PSR\, J1048$-$5832 we derived V$>$28. Moreover, we could not find evidence of counterparts of the X-ray PWNe detected by {\em Chandra} (see, Mignani et al.\ 2011b).
 
\section{DISCUSSION}

We compared our $V$-band l flux upper limits with the pulsars' rotational energy loss rates.  For PSR\, J1357$-$6429, our upper limit of $V\sim 27$ corresponds to an optical luminosity upper limit $L_{opt} \sim 0.6$--$ 21.6 \times 10^{29}$ erg~s$^{-1}$,  for a distance $d=2.4 \pm 0.6$ kpc and for an interstellar extinction $A_V = 2.2^{+1.7}_{-1.1}$,   based upon the $N_H$ derived from the fit to the X-ray spectra.. This implies  an emission efficiency upper limit $\eta_{opt} \sim 0.2$--$7 \times 10^{-7}$.  This value is at least a factor of  5 lower than the Crab pulsar and, possibly, closer to that of the Vela pulsar.  For PSR\, J1048$-$5832 our upper limit of $V\sim 27.6$ implies (for $d=2.7\pm 0.35$ kpc and $A_V = 5^{+2.2}_{-1.1}$) upper limits of  $L_{opt} \sim 0.4$--$12.5 \times 10^{30}$ erg~s$^{-1}$ 
 and $\eta_{opt} \sim 1.8$--$62.5 \times 10^{-7}$.  In principle, this does not rule out  optical emission efficiency comparable to that of the Crab pulsar, although the pulsar spin-down age (20.3 kyr) might also suggest in this case,  Vela-like emission efficiency.  
 
We also compared the  flux upper limits with the extrapolations in the optical domain of the X and $\gamma$-ray spectra of  PSR\, J1357$-$6429 and PSR\, J1048$-$5832 .  For PSR\, J1357$-$6429, we assumed an X-ray power law (PL) with photon index $\Gamma_X=1.4\pm 0.5$ plus a blackbody (BB) with temperature $kT=0.16^{+0.09}_{-0.04}$ keV ($N_H=0.4^{+0.3}_{-0.2} \times 10^{22}$ cm$^{-2}$; Esposito et al.\ 2007), 
and a $\gamma$-ray PL with photon index $\Gamma_{\gamma} =1.54 \pm 0.41$, and exponential cut-off at $\sim$ 0.8 GeV (Lemoine-Goumard et al.\  (2011),  while for PSR\, J1048$-$5832, we assumed  a PL with photon index $\Gamma_X= 2.4 \pm 0.5$ ($N_H=0.9^{+0.4}_{-0.2} \times 10^{22}$ cm$^{-2}$; Marelli et al.\ (2011),  and a  $\gamma$-ray PL with photon index $\Gamma_{\gamma}=1.38\pm 0.13$ and exponential cut-off at $\sim$ 2.3 GeV (Abdo et al.\  2009). 

The multi-wavelength spectral energy distributions (SEDs) of the two pulsars are shown in Fig. 2. For PSR\, J1357$-$6429 (Fig.2, left) we see that the optical flux upper limits can be compatible with the extrapolation of the X-ray power-law (PL). The optical flux upper limits are also compatible, with the possible exception of the R-band one, with the extrapolation of the  $\gamma$-ray PL, which does not allow us to prove that there is a break in the optical/$\gamma$-ray spectrum, as observed in other pulsars.  The  multi-wavelength SED is different  in the case of PSR\, J1048$-$5832 (Fig. 2, right), for which the optical V-band upper limit  is compatible with the extrapolation of the  steep  X-ray PL  only for the highest values of the $N_H$, but is well above the extrapolation of the   flat $\gamma$-ray one.   Interestingly enough, at variance with PSR\, J1357$-$6429, no single model can describe the optical--to--$\gamma$-ray magnetospheric emission of PSR\, J1048$-$5832. Thus, the comparison between the  multi-wavelength  SEDs of these two Vela-like pulsars  suggests that the occurrence of spectral breaks might not correlate with the pulsar's age. 

\begin{figure*}
\includegraphics[height=5.5cm,angle=0,clip=]{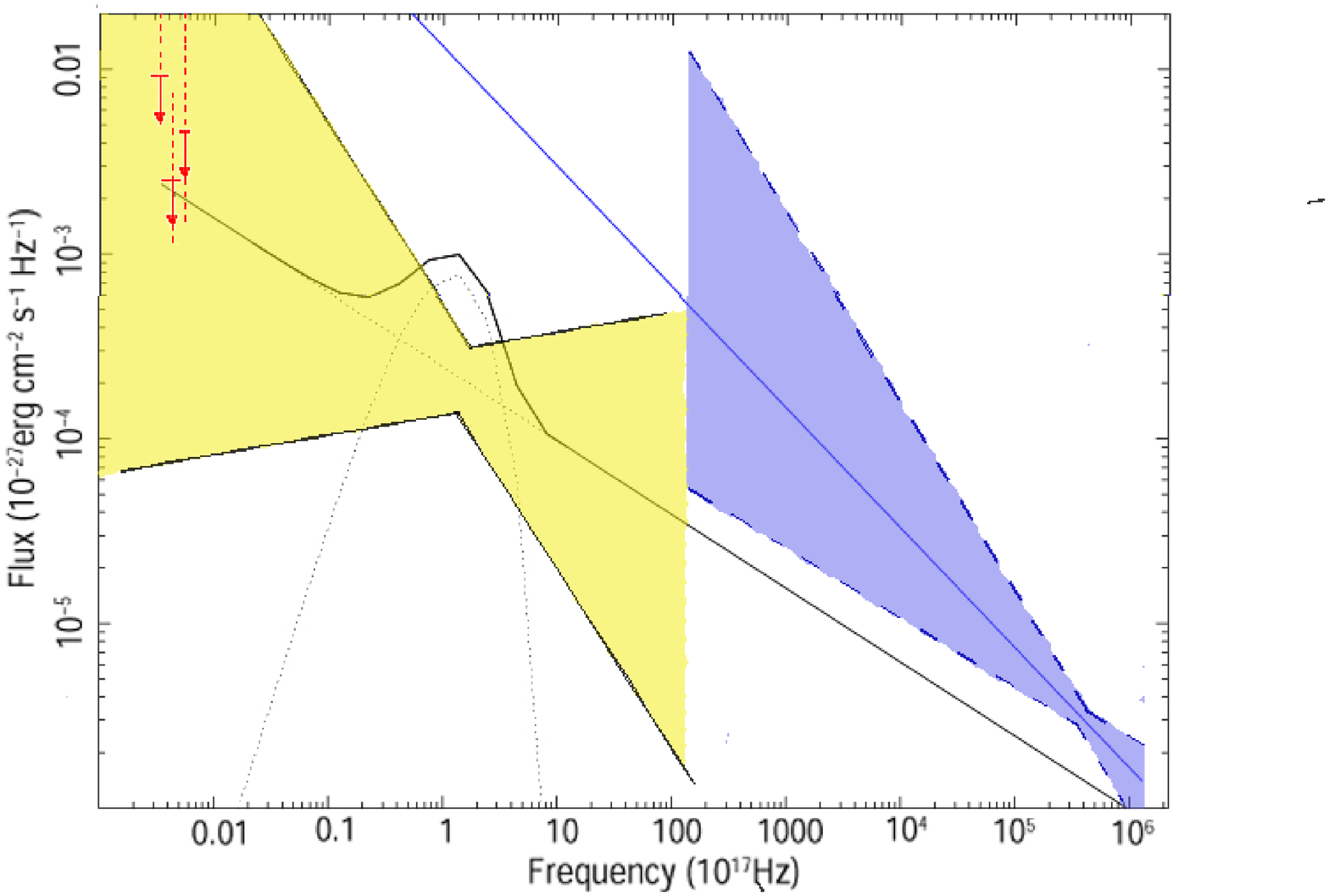}
\includegraphics[height=5.5cm,angle=0,clip=]{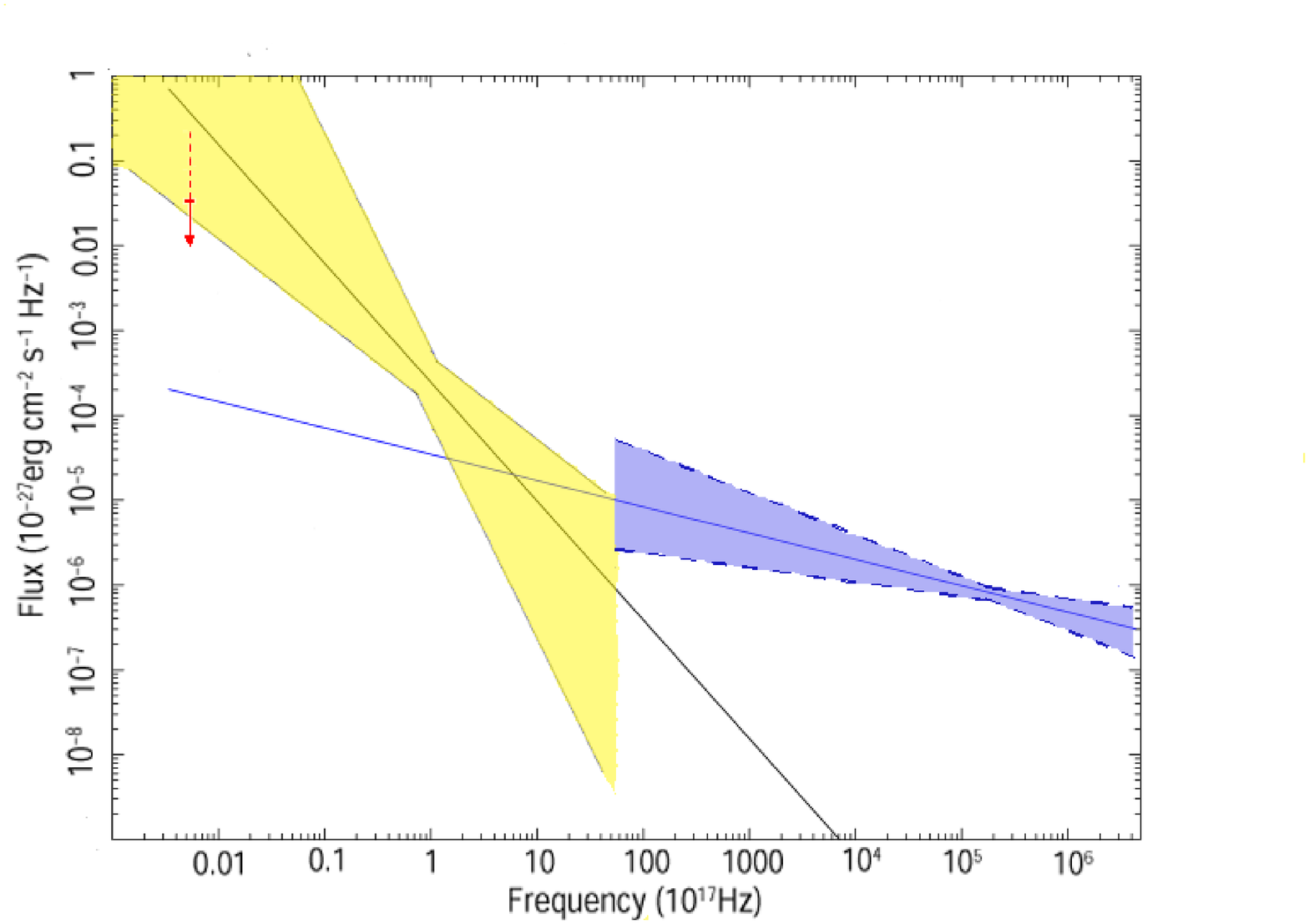}
\caption{Upper limits on the extinction-corrected optical fluxes of PSR\, J1357$-$6429 (left) and  PSR\, J1048$-$5832 (right)  compared with the low-energy extrapolations of the X-ray (solid black) and $\gamma$-ray (solid blue) spectral models that best fit the \xmm, \chan, and {\em Fermi} data.   The dotted  lines in the left panel correspond to the PL and BB components to the model X-ray spectrum.  In both panels, the yellow and blue-shaded areas  indicate the $1 \sigma$ uncertainty on the extrapolations of the X and $\gamma$-ray PLs, respectively.  The vertical red dashed lines mark the uncertainty on the extinction-corrected optical flux upper limits, computed around the best-fit value of the $N_H$. }
 \end{figure*}

\section{Summary and conclusions}

We used archival \vlt/\fors\ observations to perform the first deep optical investigations of the two  {\em Fermi} pulsars PSR\, J1357$-$6429 and  PSR\, J1048$-$5832.  We re-assessed the positions of the two pulsars from the analyses of all the available \chan\ observations and a comparison with the published radio positions. For PSR\, J1357$-$6429, this yielded a tentative proper motion of $\mu = 0.17 \pm 0.055$ arcsec yr$^{-1}$  ($70^{\circ} \pm 15^{\circ}$ position angle), which needs to be confirmed by future radio-interferometry observations. For  both pulsars, none of the objects detected around the \chan\  positions can be considered viable candidate counterparts on the basis of their relatively large optical flux, $\gtrsim 3\sigma$ offset from the pulsar position,  and lack of peculiar colours with respect to the field stars.  We then determined  $3 \sigma$ upper limits  of  $V \sim 27$ and  $V \sim 27.6$ for PSR\, J1357$-$6429 and PSR\, J1048$-$5832, respectively,  corresponding to optical emission efficiencies  $\eta_{opt} \lesssim 7\times 10^{-7}$ and $\eta_{opt} \lesssim 6 \times 10^{-6}$.
The \vlt\ observations presented here are close to the sensitivity limits achievable with 10m-class telescopes under sub-arcsec seeing conditions. In the case of PSR\, J1357$-$6429, deep, high spatial resolution images with the \hst\ are probably required to firmly claim an object detection from the flux enhancement seen at the \chan\ position (see section 3). In the case of PSR\, J1048$-$5832,  the large interstellar extinction ($A_V \sim 5$) 
hamper observations in the optical/near-ultraviolet. Deep observations in the near-infrared, either with the \hst\ or with adaptive optic device at 10m-class telescopes, might represent a better opportunity to spot a candidate counterpart to the pulsar.

\bigskip 
\begin{acknowledgments}
The authors thank the {\em Fermi} Pulsar Timing Consortium, in particular David Smith, Ryan Shannon, and Simon Johnston, for checking for updated radio coordinates of the two pulsars. 
\end{acknowledgments}

\end{document}